\documentclass{ws-procs9x6}

\usepackage{wrapfig}
\usepackage{multirow}
\usepackage{subfigure}

\newcommand{\Jpsi}{\ensuremath{J/\psi\,}}
\newcommand{\Ups}{\ensuremath{\Upsilon\,}}
\newcommand{\GeV}{\ifmmode {\mathrm{\ Ge\kern -0.08em V}}\else
                   \textrm{Ge\kern -0.08em V}\fi}%
\newcommand{\MeV}{\ifmmode {\mathrm{\ Me\kern -0.08em V}}\else
                   \textrm{Me\kern -0.08em V}\fi}%

\begin{document}

\title{Quarkonium production and polarisation with early data at ATLAS}

\author{DARREN D.~PRICE\footnote{on behalf of the \uppercase{A}tlas \uppercase{C}ollaboration.}}

\address{Department of Physics, Lancaster University, \\
Lancaster, LA1 4YB, UK \\
E-mail: Darren.Price@cern.ch}

\maketitle

\abstracts{
One of the first physics results to come out of ATLAS will be an analysis of $J/\psi$ and $\Upsilon$ production at 14~TeV.
I give an overview of the motivation for looking at the theoretical model underlying quarkonium production,
ATLAS expected performance and rates for quarkonium reconstruction and ability to separate out various
proposed production models with a view to improving our understanding of QCD.
}

\section{Introduction and motivation}

When switched on, the LHC will produce charm and beauty quarks in abundance even in
low luminosity runs during the first few years of running and the production rate of 
quarkonia such as \Jpsi and \Ups, important for many physics studies, will also be large. 
The reasonable branching fraction of both the \Jpsi and the
\Ups into charged lepton pairs allows for easy separation of these events from
the huge hadronic background at the LHC.

Their importance for ATLAS is threefold: first, being narrow resonances, they can be 
used as tools for alignment and calibration of the trigger, tracking and muon systems. 
Secondly, understanding the details of the prompt onia production is a challenging task
and a good testbed for various QCD calculations, spanning both perturbative and non-perturbative
regimes. Last, but not the least, heavy quarkonia are among the decay products of 
heavier states, serving as good signatures in searches for rare decays and CP-violating processes
which form the backbone of the long-term ATLAS B-Physics programme.
These processes have prompt quarkonia as a background and, as such, a good description of 
the underlying quarkonium production process is crucial to the success of these studies.

\section{Overview and current status}

The Colour Singlet Model (CSM)\,\cite{CSM} of quarkonium production enjoyed some success before CDF
measured an excess of direct \Jpsi production\,\cite{Abe:CDF} more than an order of magnitude
greater than predicted, with incorrect $p_T$ dependence.
The Colour Octet Mechanism (COM)\,\cite{COM} was proposed as a solution to this problem, 
suggesting that the heavy quark pairs could evolve into a quarkonium state with particular
quantum numbers through radiation of soft gluons during hadronisation. 

Despite the successes of COM\,\cite{kramer-cdf}, recent measurements at CDF\,\cite{Abulencia:2007us} 
and D\O\,\cite{D0polar} show disagreement with COM predictions on polarisation. 
The polarisation of the quarkonium state can be determined
by measuring the parameter $\alpha=(\sigma_T-2\sigma_L)/(\sigma_T+2\sigma_L)$, 
which may vary between $+1$ for 100\% transversely polarised, 
to $-1$ for 100\% longitudinally polarised production. This can be 
achieved by measuring the distribution of $\theta^\ast$, the angle between the
positive muon from the quarkonium decay (in the quarkonium rest frame) and the quarkonium
direction in the lab frame. Various production models\,\cite{COM,othermodels} predict 
different $p_T$ dependencies of the quarkonium polarisation, so this quantity serves as an
important measurement for discriminating these models. 

It is interesting to note that the results from D\O\, and CDF
disagree both with each other, and various theoretical predictions. Both experiments 
suffer from low acceptance in the discriminating high $|\cos\theta^\ast|$ region.
ATLAS is expected to be capable of detailed checks of the predictions of these and other
models in a wider range of $|\cos\theta^\ast|$, $p_T$ and $\eta$.

\section{Observation of quarkonium in ATLAS}

Two main trigger scenarios are considered here for the study of prompt quarkonia. The first is a di-muon trigger
which requires two identified muons, both with a $4$~\GeV~$p_T$ threshold and within a pseudorapidity 
of $|\eta|<2.4$. The di-muon sample considered here has offline cuts of $6$~and $4$~\GeV~$p_T$ applied
to the two identified muons. The second scenario is a single muon trigger in which only 
one identified muon is required, with a $p_T>10$~\GeV and $|\eta|<2.4$, which is combined offline
with Inner Detector tracks reconstructed down to a minimum $p_T$ of $0.5$~\GeV.

\begin{figure}[htb]
  \begin{center}
    \includegraphics[width=9cm]{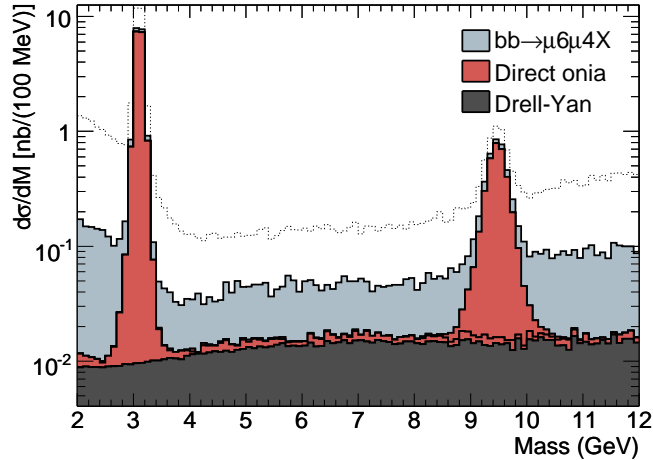}
    \caption{Sources of low invariant mass di-muons, after background suppression cuts. 
      The light dotted line shows the background level before vertexing and proper-time cuts.
    }
    \label{fig:lowmass_withcuts}
  \end{center}
\end{figure}

The mass resolutions at ATLAS are expected to be approximately
$54$~\MeV (\Jpsi) and $170$~\MeV (\Ups). 
The main expected sources of background are indirect \Jpsi 
from $B$-decays, the continuum of muons from heavy flavour 
decays, Drell-Yan and decays in flight of $K^\pm$ and $\pi^\pm$.
Figure~\ref{fig:lowmass_withcuts} shows the reconstructed
invariant mass distribution in the \Jpsi and \Ups region for the di-muon dataset.
For an integrated luminosity of 10~pb$^{-1}$ we expect signal yields to be approximately
150,000 \Jpsi and 25,000 \Ups. The background under the \Jpsi and \Ups peaks is
suppressed with vertexing and impact parameter cuts on the muons and a pseudo-proper time
cut on the reconstructed quarkonium candidate.  

\section{Polarisation and cross-section measurement}
\label{sec:polarisation}

It can be seen that $\cos\theta^\ast\simeq 0$ corresponds to events where both
muons have roughly equal transverse momenta, while in order to have $\cos\theta^\ast$ 
close to $\pm 1$ (the crucial area) one muon's $p_T$ needs to be very high with the other 
$p_T$ very low. In the case of a di-muon trigger, both muons from the \Jpsi and 
\Ups decays are forced to have relatively large transverse momenta.
Whilst this condition allows both muons to be identified, it also severely restricts
acceptance in the polarisation angle $\cos\theta^\ast$, meaning that for a given $p_T$ 
a significant fraction of the total cross-section is lost. 

\begin{wrapfigure}{l}{5.5cm}
  \begin{center}
    \centering
    \includegraphics[width=5.5cm]{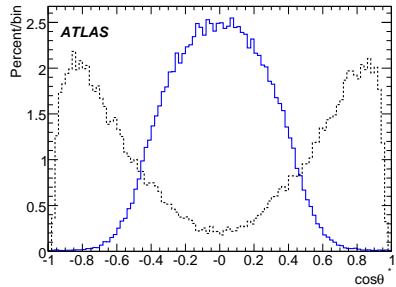}
    \caption{Acceptance of $\cos\theta^\ast$ in di-muon (solid line) and single muon (dashed line) 
      datasets for \Jpsi. The {\em true} angular distribution in both cases is flat.
    }
    \label{fig:acceptancepolarisation-jpsi}
  \end{center}
\end{wrapfigure}

With the single muon trigger one removes the constraint on the second muon. Now one has a
high $p_T$ muon which triggered the event and one reconstructed track down to a
$p_T$ threshold around 0.5~\GeV. Thus, the onium events with a single muon
trigger typically have high values of $|\cos\theta^\ast|$, complementing the
di-muon sample (see Figure~\ref{fig:acceptancepolarisation-jpsi}, which shows the
acceptance for both triggers).
Combined carefully together the single and di-muon samples provide excellent
coverage across almost the entire range of $\cos\theta^\ast$ over the same onia
$p_T$ range. The measured distributions from the combined data sample are corrected for acceptance and
efficiencies to recover the true underlying distribution.

\begin{figure}[htb]
  \begin{center}
    \centering
      \includegraphics[width=9cm]{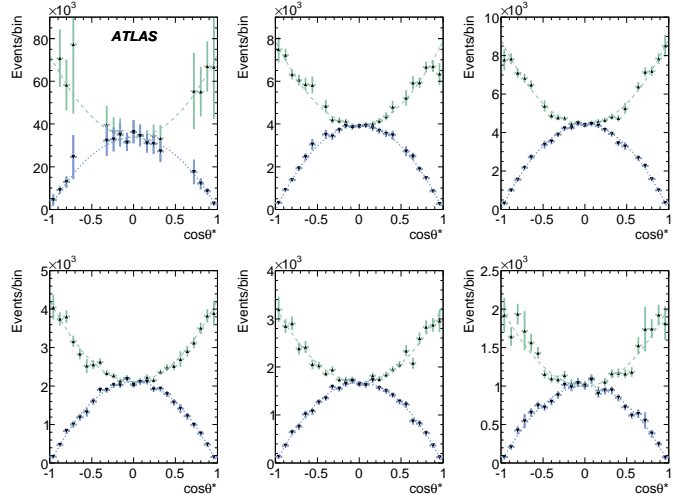}
      \caption{Acceptance and efficiency corrected \Jpsi polarisation angle distribution 
	($\alpha_{gen}=\pm 1$) for increasing $p_T$ slices (see Table~\ref{tab:measurement}). 
	Statistics for $\int\mathcal{L}=10$~pb$^{-1}$.
      }
      \label{fig:correctedpolarisation-jpsi}
  \end{center}
\vspace*{-13pt}
\end{figure}

Three MC datasets were produced, with fully transverse/ longitudinal
polarisation and one with zero polarisation, to test the ability of ATLAS in these
limit cases.
The resultant distributions for the $\alpha_{gen}=\pm 1$ datasets are shown
in Figure~\ref{fig:correctedpolarisation-jpsi} and are detailed for all three in
Table~\ref{tab:measurement}, along with the corrected cross-section in that $p_T$ bin.
The errors shown include statistical errors on the data,
as well as the systematic effects of uncertainties on the acceptances and efficiencies.

\begin{table}[hbt]
\tbl{\Jpsi and \Ups polarisation and cross-sections measured in slices of $p_T$,
  for 10~pb$^{-1}$.}
{\footnotesize
\begin{tabular}{ccrrrrrr}
\hline
 Sample & $p_T$, \GeV & $9-12$ & $12-13$ & $13-15$ & $15-17$ & $17-21$ & $>21$ \\
\hline
\hline

\multirow{2}{*}{\Jpsi} &
$\alpha$ & $0.156$ & $-0.006$ & $0.004$ & $-0.003$ & $-0.039$ & $0.019$ \\
& & $\pm0.166$ & $\pm0.032$ & $\pm0.029$ & $\pm 0.037$ & $\pm 0.038$ & $\pm 0.057$ \\
\cline{2-8}
\multirow{2}{*}{$\alpha_{\mathrm{gen}}=0$} &
$\sigma$, nb & $87.45$ & $9.85$ & $11.02$ & $5.29$ & $4.15$ & $2.52$ \\
& & $\pm4.35$ & $\pm0.09$ & $\pm0.09$ & $\pm 0.05$ & $\pm 0.04$ & $\pm 0.04$ \\
\hline
\hline

\multirow{2}{*}{\Jpsi} &
$\alpha$ & $1.268$ & $0.998$ & $1.008$ & $0.9964$ & $0.9320$ & $1.0217 $ \\
& & $\pm0.290$ & $\pm0.049$ & $\pm0.044$ & $\pm 0.054$ & $\pm 0.056$ & $\pm 0.088$ \\
\cline{2-8}
\multirow{2}{*}{$\alpha_{\mathrm{gen}}=+1$} &
$\sigma$, nb & $117.96$ & $13.14$ & $14.71$ & $7.06$ & $5.52$ & $3.36$ \\
& & $\pm6.51$ & $\pm0.12$ & $\pm0.12$ & $\pm 0.07$ & $\pm 0.05$ & $\pm 0.05$ \\
\hline
\hline

\multirow{2}{*}{\Jpsi} &
$\alpha$ & $-0.978$ & $-1.003$ & $-1.000$ & $-1.001$ & $-1.007$ & $-0.996 $ \\
& & $\pm0.027$ & $\pm0.010$ & $\pm0.010$ & $\pm 0.013$ & $\pm 0.014$ & $\pm 0.018$ \\
\cline{2-8}
\multirow{2}{*}{$\alpha_{\mathrm{gen}}=-1$} &
$\sigma$, nb & $56.74$ & $6.58$ & $7.34$ & $3.53$ & $2.78$ & $1.68$ \\
& & $\pm2.58$ & $\pm0.06$ & $\pm0.06$ & $\pm 0.04$ & $\pm 0.03$ & $\pm 0.02$ \\
\hline
\hline

\multirow{2}{*}{\Ups} &
$\alpha$ & $-0.42$ & $-0.38$ & $-0.20$ & $0.08$ & $-0.15$ & $0.47 $ \\
& & $\pm0.17$ & $\pm0.22$ & $\pm0.20$ & $\pm 0.22$ & $\pm 0.18$ & $\pm 0.22$ \\
\cline{2-8}
\multirow{2}{*}{$\alpha_{\mathrm{gen}}=0$} &
$\sigma$, nb & $2.523$ & $0.444$ & $0.584$ & $0.330$ & $0.329$ & $0.284$ \\
& & $\pm0.127$ & $\pm0.027$ & $\pm0.029$ & $\pm 0.016$ & $\pm 0.015$ & $\pm 0.012$ \\
\hline
\end{tabular}\label{tab:measurement} }
\vspace*{-7pt}
\end{table}

With an integrated luminosity of just $10$~pb$^{-1}$, due to the high rate 
ATLAS aims to measure the polarisation of prompt vector quarkonium states
to far higher transverse momenta than previous experiments with extended
coverage in $\cos\theta^\ast$, which will allow for improved
fidelity of efficiency measurements and thus reduced systematics.
The precision of the \Jpsi polarisation measurement can reach 
$0.02-0.06$ (dependent on the level of polarisation itself).


\end{document}